# A coarse-grained model for the elastic properties of cross linked short carbon nanotube/polymer composites

Atiyeh Alsadat Mousavi[c,∗], Behrouz Arash[c], Xiaoying Zhuang[d,c], Timon Rabczuk[a,b,e,c,∗]

[a]*Division of Computational Mechanics, Ton Duc Thang University, Ho Chi Minh City, Vietnam.*
[b]*Faculty of Civil Engineering, Ton Duc Thang University, Ho Chi Minh City, Vietnam.*
[c]*Institute of Structural Mechanics, Bauhaus Universität-Weimar, Marienstr 15, D-99423 Weimar, Germany*
[d]*Department of Geotechnical Engineering, Tongji University, Shanghai, China*
[e]*School of Civil, Environmental and Architectural Engineering, Korea University, Seoul, South Korea*

**Abstract**

Short fiber reinforced polymer composites have found extensive industrial and engineering applications owing to their unique combination of low cost, relatively easy processing and superior mechanical properties compared to their parent polymers. In this study, a coarse-grained (CG) model of cross linked carbon nanotube (CNT) reinforced polymer matrix composites is developed. A characteristic feature of the CG model is the ability to capture the covalent interactions between polymer chains, and nanotubes and polymer matrix. The dependence of the elastic properties of the composites on the mole fraction of cross links, and the weight fraction and distribution of nanotube reinforcements is discussed. The simulation results reveal that the functionalization of CNTs using methylene cross links is a key factor toward significantly increasing the elastic properties of randomly distributed short CNT reinforced poly (methyl methacrylate) (PMMA) matrix. The applicability of the CG model in predicting the elastic properties of CNT/polymer composites is also evaluated through a verification process with a micromechanical model for unidirectional short fibers.

## 1. Introduction

Short fiber reinforced polymer (SFRP) composites have attracted intense attention due to their ease of fabrication, low manufacturing costs and superior mechanical, thermal and electrical properties [1, 2, 3]. SFRP composites achieve high levels of stiffness comparable to continuous fiber reinforced polymer composites. Simultaneously, the flexibility of unreinforced polymers to be formed into complex shapes, which is suitable in automotive, aerospace and chemical industries, is preserved [4, 5]. Among various types of fibers used in the composites, carbon nanotubes (CNT) are promising as ultra-high-strength reinforcements because of their remarkable mechanical properties [6].

∗Corresponding authors
*Email addresses:* `atiyeh.alsadat.mousavi@uni-weimar.de` (Atiyeh Alsadat Mousavi),
`timon.rabczuk@tdt.edu.vn` (Timon Rabczuk)

In order to understand the mechanical behavior of short CNT/polymer composites, a range of studies have been conducted using molecular dynamics (MD) simulations [7, 8, 9, 10]. Zhu et al. [11] studied the elastic properties of an epoxy Epon 862 matrix with a size of $4.028 \times 4.028 \times 6.109$ $nm^3$ reinforced by short (10, 10) CNTs with length-to-diameter aspect ratios of 2.15 and 4.5. They indicated that a unidirectional short CNT reinforcement with a aspect ratio of 4.5 increases the Young's modulus of the composite up to 20% compared to the pure Epon 862 matrix. MD studies on the elastic properties of short single-walled CNT (SWCNT) reinforced Poly (vinylidene fluoride) (PVDF) matrix composites [12] showed that a (5, 5) SWCNT with a length of 2 nm can increase the Young's modulus of a CNT/PVDF composite by 1 GPa. The simulation unit cell consists of a (5, 5) SWCNT with a volume fraction of 1.6% embedded in 60 PVDF chains. Arash et al. [13] investigated the mechanical behavior of CNT/poly (methyl methacrylate) (PMMA) polymer composites under tension. They proposed a method to evaluate the elastic properties of the interfacial region of CNT/polymer composites. The CNT/polymer composite is simulated to obtain the elastic properties of a PMMA polymer matrix with a size of $3.7 \times 3.7 \times 8$ $nm^3$ reinforced by a short (5, 5) SWCNT. Their simulation results reveal that the Young's modulus of the composite increases from 3.9 to 6.85 GPa with an increase in the length-to-diameter aspect ratio of the nanotube from 7.23 to 22.05.

The mechanical properties of reinforced polymer composites strongly depends on the strength of interactions between polymer chains and CNTs, which in turn affects the performance of load transfer between polymer matrix and nanotune reinforcements. Two major methods proposed in the literature to enhance the mechanical properties of the nanocomposites are: (1) the application of helical polymer chains wrapping around nanotubes to increase the adhesion strength between CNTs and polymer chains [14] , and (2) the formation of covalent cross links between nanotube reinforcements and polymer matrix for strengthening the interface between nanotubes and polymer matrices [15, 16, 17]. Frankland et al. [15] studied the effects of two methylene unit ($2CH_2$) cross links between polymer chains and carbon nanotubes on the elastic properties of CNT/polymer composites. They modeled a (10, 10)

CNT embedded in a polyethylen matrix using molecular dynamics (MD) simulations, and showed that even a relatively low density of the cross links can have a considerable influence on the elastic properties of the composites. Min et al. [17] investigated the shear response of PMMA polymer cross linked by ethylene glycol dimethyl acrylate (EGDMA) using molecular simulations. It was reported that a cross link density of 1.15% significantly affects the stress response of the polymer material and the cross linked polymer exhibits a more ductile behavior compared to its linear counterpart.

Although MD simulations have been broadly utilized in modelling reinforced polymer nanocomposites, the immense computational cost required by the simulations severely limits their applicability to small molecular systems over a limited time scale. This drawback make the MD simulations unable to study the effect of fiber sizes and distributions on the mechanical behaviour of reinforced polymer composites. To overcome the MD limitations, coarse-grained (CG) models that span from nanoscale to mesoscale have been introduced in the literature [18, 19, 20]. In CG models, a set of atoms are mapped to a CG bead. A CG bead would not only extend the accessible time and length-scales but also enables to partially maintain molecular details of an atomistic system. Up to now, many polymer materials have been simulated by CG models [19, 21, 22]. Recently, the reliability of CG models has been tested in modelling of graphenes and CNTs [23, 24, 25, 26]. A CG model has been introduced for the elastic and fracture behavior of graphenes with a $\sim 200$ fold which can increase the computational speed compared to full atomistic simulations [25]. Zhao et al. [26] calibrated parameters of the CG stretching, bending and torsion potentials for SWCNTs in order to consider their static and dynamic behaviours. Parameters of non-bonded van der Waals (vdW) interactions between CNTs in a bundle were obtained. They established a CG model with a potential for analysing the mechanical properties of CNT bundles while decreasing the computational costs compared to atomistic simulations. Arash et al. [27] developed a comprehensive CG model of polymer composites reinforced by carbon nanotubes. The proposed model was able to obtain the non-bonded interactions between polymer chains and nanotubes. They then used the model to study the elastic properties of short CNT/PMMA polymer composites.

Despite the CG simulation studies on the elastic properties of randomly distributed short CNT reinforced polymer composites, there is still no CG simulation investigations on the mechanical properties of the composites with covalent cross links. The effects of cross links between polymer chains, and nanotubes and polymer chains on the elastic properties of ran-domly distributed CNTs reinforced polymer matrix have been not efficiently understood. Hence, a quantitative study on the elastic properties of the composites is essential to achieve a comprehensive understanding of their mechanical characteristics.

This study aims to develop a CG model of cross linked CNT/PMMA composites to investigate their mechanical behavior in the elastic regime. The CG force field parameters for EGDMA cross links between polymer chains, and $2CH_2$ cross links between CNTs and polymer matrix are calibrated using results obtained from molecular simulations. The effects of cross links between polymer chains, and nanotube and polymer matrix on the elastic properties of randomly distributed CNT/PMMA composites are studied in detail. The proper RVE size, representing the whole microstructure of randomly distributed CNT reinforced polymer composites, is explored. The effects of weight fractions and distribution of CNTs on the elastic properties of the nanocomposites are examined. The applicability of the CG model to obtain the elastic properties of unidirectional CNT/PMMA composites is also interpreted using a micromechanical model.

## 2. Methodology

In this study, a CG model that was previously proposed [27] is used to simulate CNT/PMMA composites. In the CG model used in this paper, each monomer of methyl methacrylate ($C_5O_2H_8$) is mapped into a CG bead hereafter named P bead with an atomic mass of 100.12 amu as illustrated in Fig.1a. The center of the bead is chosen to be the center of mass of the monomer. Each five atomic rings of (5, 5) CNTs is treated as a CG bead with an atomic mass of 600.55 amu defined by C bead as shown in Fig.1b. The center of C beads is assumed to be the center of the five atomic rings. In the CG model, compared to their full atomistic systems, the degrees of freedom (DOF) decrease to 15 and 50 folds for P and C beads, respectively.

The total potential energy can be written as,

$$E_{total}(d, \theta, r) = \sum_i E_{b_i} + \sum_j E_{a_j} + \sum_{lm} E_{vdW_{lm}} + E_0, \qquad (1)$$

where $E_b$, $E_a$ and $E_{vdW}$ are the terms of energy corresponding to the variation of the bond length, the bond angle and the van der Waals (vdW) interaction, respectively. In Eq. (1), $E_0$ corresponds to the constant free energy of the system. The functional forms of $E_b$ and $E_a$, associated with a single interaction, are

$$E_b(d) = \frac{k_d}{2}(d - d_0)^2 \quad for \ d < d_{cut}, \qquad (2)$$

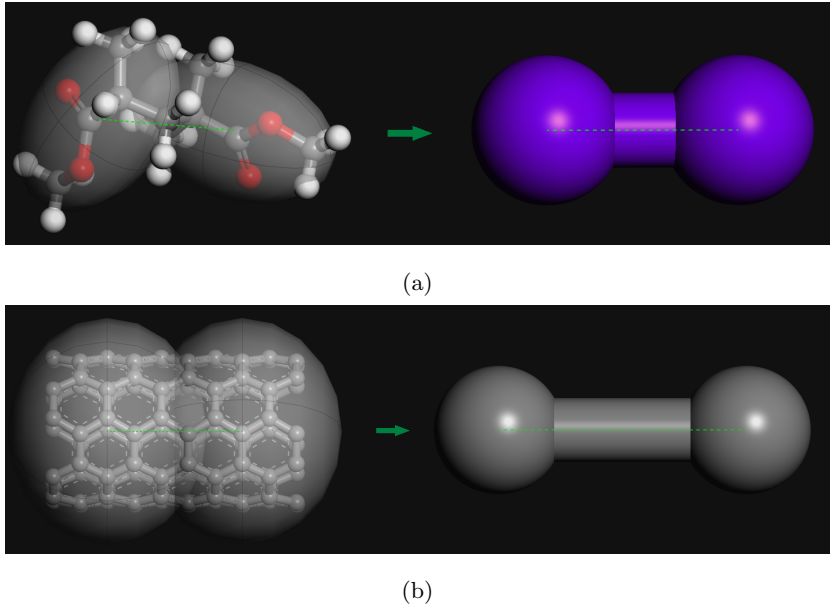

(a)

(b)

Figure 1: CG model representations resulting from (a) two monomers of a PMMA polymer chain and (b) a (5, 5) CNT with 10 rings of carbon atoms.

and

$$E_a(\theta) = \frac{k_\theta}{2}(\theta - \theta_0)^2. \tag{3}$$

In Eq. (2), $k_d$ is the spring constant of the bond length and $d_0$ is the equilibrium bond distance. In Eq. (3), $k_\theta$ and $\theta_0$ represent the spring constant and the equilibrium bond angle, respectively. Finally, the functional form of the third term of the total energy is obtained by the most common expression of Lennard-Jones potential

$$E_{vdW}(r) = D_0[(\frac{r_0}{r})^{12} - 2(\frac{r_0}{r})^6], \tag{4}$$

where $D_0$ and $r_0$ are associated with the equilibrium well depth and the equilibrium distance, respectively. The cutoff distance which can be calculated by vdW interactions, is set to be 1.25 nm. In Table 1, the CG force fields parameters are represented [27].

As discussed, an effective way to enhance the elastic properties of carbon nanotube reinforced polymer composites is the formation of covalent cross links. Herein, we respectively choose $2CH_2$ and EGDMA cross links between polymer matrix and CNTs [15, 16], and polymer chains [17]. The atomistic and CG models of a $2CH_2$ between a polymer chain and a CNT are illustrated in Figs.2a and 2b, respectively. The atomistic model and its corresponding CG counterpart of a EGDMA cross link between two PMMA polymer chains are also shown in Figs.2c and 2d . The cross links are randomly added nanotubes and polymer matrix, and be-tween PMMA chains according to the following rules: (1) cross links can only connect beads that are placed within the equilibrium distance, (2) no cross link connects a chain to itself, and (3) there are at least two beads between two sequential cross links. The parameterization of CG stretching potentials for the $2CH_2$ and the EGDMA cross links are described below.

Table 1: Parameters of the CG force field for C, P beads.

| Type of interaction | Parameters | C bead | P bead | C-P beads |
|---|---|---|---|---|
| Bond | $K_0$ ($kcal/mol/Å^2$) | 1610.29 | 194.61 | — |
| | $d_0$ (Å) | 5.95 | 4.05 | — |
| Angle | $K_\theta$ ($kcal/mol/Å^2$) | 64280 | 794.89 | — |
| | $\theta_0(°)$ | 180 | 84.8 | |
| vdW | $D_0$ $kcal/mol$ | 9.45 | 6.53 | 7.7 |
| | $r_0$ (Å) | 10.68 | 1.125 | 2.8 |

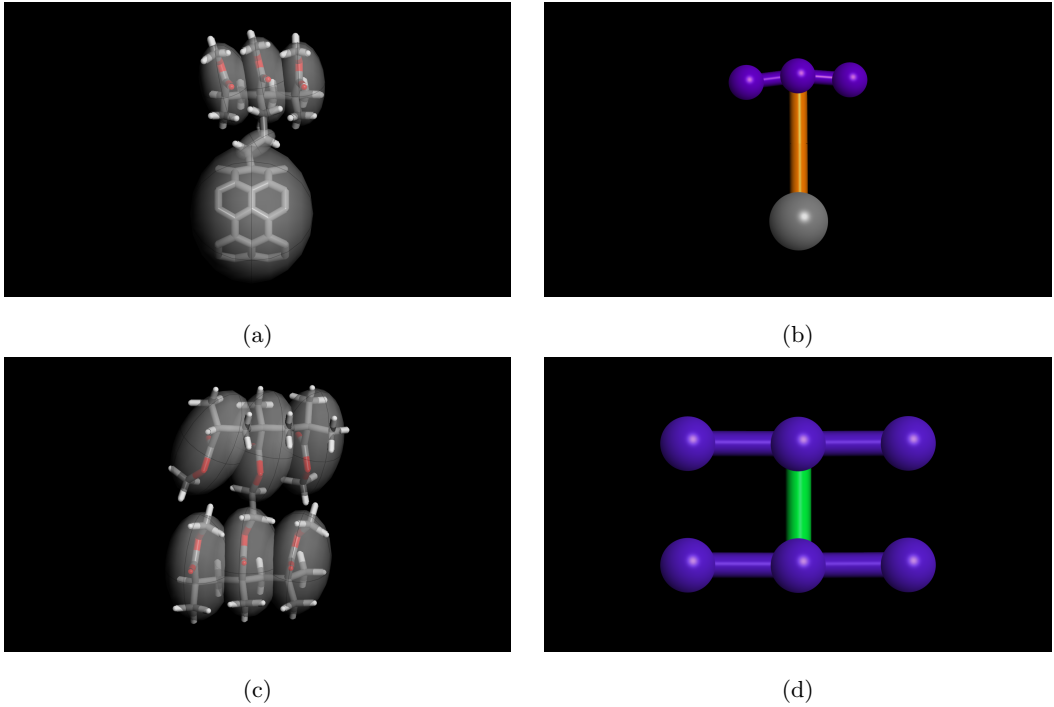

(a)     (b)

(c)     (d)

Figure 2: (a) The atomistic model and (b) its corresponding CG illustration of $2CH_2$ cross link between a PMMA monomer and a CNT. An EGDMA cross link between two PMMA monomers is shown in (c) atomistic illustration (d) and its CG model.

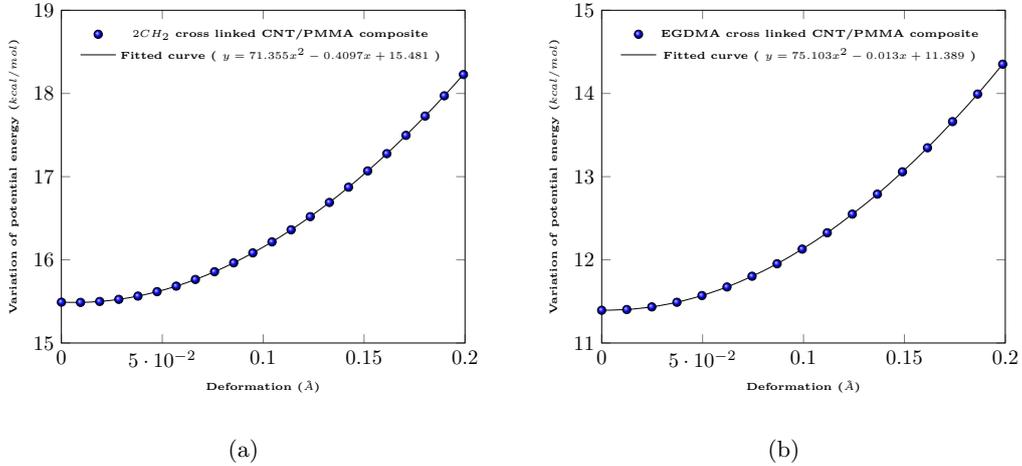

Figure 3: Variation of potential energy of the cross linked nanotube/polymer system versus deformations from which the spring constant is calculated for (a) $2CH_2$ and (b) EGDMA cross links

For small deformations along the centers of a nanotube with 5 atomic rings (defined as a C bead) and a monomer (defined as a P bead) connected by a $2CH_2$ cross link, the total potential energy, $U$, obtained by molecular simulations can be equalized to the stretching potential of a two-bead CG model as $U(d) = \frac{1}{2}K_d(d-d_0)^2$. $d0$ is the equilibrium distance between two beads measured to be $9.487(\text{Å})$ for $2CH_2$ cross links using molecular simulations. The spring constant of the bond length is then given by the second derivative of the potential energy with respect to the bond length as $k_d = \frac{\partial^2 U}{\partial d^2}$. Fig. 3a presents the variation of potential energy of the cross linked nanotube/polymer system versus deformations from which the spring constant is calculated to be $142.71(kcal/mol/\text{Å}^2)$ for $2CH_2$ cross link. Similar to previous simulations, the spring constant of the bond length is determined for EGDMA cross links between polymer chains. The equilibrium distance is also measured to be $d_0 = 6.21(\text{Å})$. The potential energy of a two monomer system connected by an EGDMA cross link under a longitudinal deformation obtained by molecular simulations is equated to the potential energy of the corresponding CG model. The variation of the potential energy of the system is shown in Fig. 3b from which the spring constant of the bond length of the CG model is calculated to be $150.206(kcal/mol/\text{Å}^2)$.

In the molecular simulations, COMPASS force field [28] is used to describe intermolecular interactions. It is the first ab initio force-field that enables an accurate prediction of the mechanical behavior of CNTs and polymers. The non-bonded interactions are modeled using the vdW and coulombic interaction energy terms. A potential cutoff of $1.2 nm$ is used in calculation of the non-bonded interactions. Partial charges of atoms are also assigned using Qeq method [29]. In this study, Accelrys Material Studio 7.0 is used for conducting the simulations.

## 3. Results and discussion

### 3.1. Verification of the CG Model

In order to examine the applicability of the CG model, we first study the elastic properties of PMMA polymer matrix reinforced by CNTs aligned in the load direction. A unit cell with the size of $12 \times 12 \times 12\ nm^3$ and periodic boundary conditions is initially constructed as illustrated in Fig. 4. The unit cell contains PMMA polymer chains with a mass density of $1.1 \frac{g}{cm^3}$ and unidirectional 10-nm long (5, 5) CNTs. Each polymer chain is composed of 60 repeated monomer units.

To find a global minimum energy configuration, a geometry optimization is first performed using the conjugate-gradient method [30]. The system is then relaxed by the isothermal-isobaric ensemble (NPT) at room temperature of 298 K and atmospheric pressure of 100 kPa for 100 ps with a time step of 1 fs. Another NPT simulation is continued at the same atmospheric pressure at room temperature of 100 kPa for 4 ns with a time step of 10 fs. The Andersen feedback thermostat [31] and the Berendsen barostat algorithm [32] are respectively used for the system temperature and pressure conversions. The NPT simulations are followed by a further energy minimization. The combination of energy minimization and dynamic simulations guarantee the removal of internal stresses in the composite system. After the preparation of the system, the constant-strain minimization method is used to calculate the elastic properties of the composite. A small tensile strain of 1% with an increment of 0.02% is applied to the CNT/PMMA composite in the x-direction where the CNTs are aligned. The potential energy of the structure is re-minimized after each increment of the applied strain. The tensile strain is accomplished by uniformly expanding the dimensions of the simulation cell in the loading direction and re-scaling the new coordinates of the atoms to fit within the new dimensions. The stress of the composite is then obtained according to the virial stress definition. During the tensile deformation, the pressure in the y and z-directions is kept at atmospheric pressure by controlling the lateral dimensions. This process provides the variation of stress versus applied strain from which the Young's modulus is obtained.

Table 2: Young's modulus of a CNT/PMMA polymer composite with different CNT weight fractions. The RVE size is $12 \times 12 \times 12$ $nm^3$ reinforced by 10-nm long (5, 5) CNTs aligned in the load direction.

| CNT Weight fraction | Present CG model | Krenchel's rule of mixtures |
|---|---|---|
| $wt\%$ | Young's modulus ($GPa$) | Young's modulus ($GPa$) |
| Pure polymer | 2.88 | - |
| 3 | 3.34 | 3.38 |
| 5 | 3.63 | 3.70 |
| 8 | 4.22 | 4.44 |
| 10 | 4.52 | 4.85 |

The effect of the CNT weight fractions on the elastic properties of the CNT/PMMA composite is presented in Table 2. The weight fraction of CNTs varies from 3 to 10 $wt\%$. To adjust the value of CNT weight fraction, the number of CNTs in the polymer matrix differs from 4 to 12. From Table 2, the Young's modulus of the CNT/PMMA composite increases from 3.34 to 3.63 $GPa$ with an increase in the CNT weight fraction from 3 to 5 $wt\%$, respectively showing a percentage increase from 16 to 26% compared to the pure polymer. The Young's modulus further increases to 4.22 and 4.52 $GPa$ for the CNT weight fraction of 8 and 10 $wt\%$, revealing percentage increases of 46 and 57%, respectively. The simulation results indicate a slight increase in the Young's modulus of PMMA matrix reinforced by short unidirectional CNTs at the weight fraction of 3 $wt\%$, while the CNT reinforcements with a weight fraction of 10 $wt\%$ significantly enhance the stiffness of the CNT/PMMA composite.

The elastic properties of the CNT/PMMA composite with unidirectional CNTs, simulated by the CG model, can be interpreted by micromechanical continuum models. Herein, we use the Krenchel's rule of mixtures to calculate the Young's modulus of the CNT/PMMA composite as [33],

$$E^c = \eta_0 \eta_l E^f V^f + E^m V^m, \tag{5}$$

where superindices c, f and m represent the composite, fiber and matrix, respectively. $E$ and $V$ are the Young's modulus and volume fraction of the materials. The Young's modulus of a (5, 5) CNT reinforcement is measured to be 1.65 TPa using MD simulations [13]. It is noteworthy that in calculation of the Young's modulus of the CNT, the nanotube is supposed to be a solid bar with a cross sectional area of $A_{CNT} = \frac{\pi d^2}{4}$. The terms $\eta_0$ and $\eta_l$ illustrate the efficiency factors of fiber length and orientation. $\eta_0$ is equal to 1 in the case of unidirectional fibers while $\eta_l$ is calculated by [34],

$$\eta_l = 1 - \frac{tanh\frac{\zeta L^f}{2}}{\frac{\zeta L^f}{2}}. \tag{6}$$

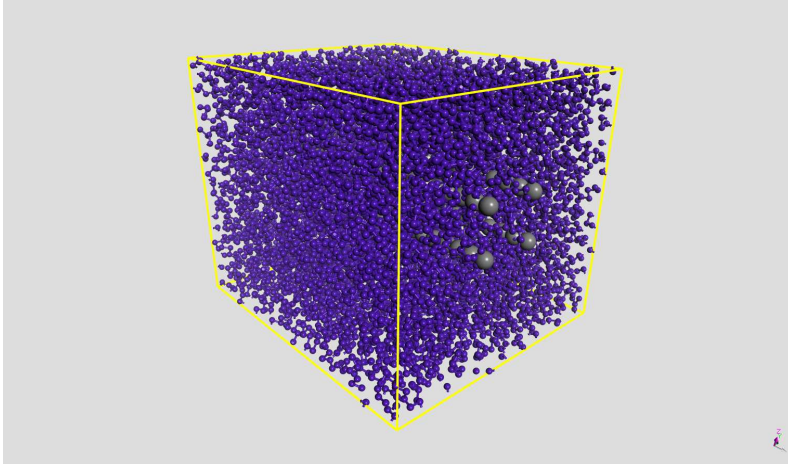

Figure 4: Initial configuration of a PMMA matrix reinforced by 10-nm long (5, 5) CNTs aligned in the load direction. The weight fraction is set to be 3 $wt\%$.

The term $\zeta$ in Eq. (6) is given by

$$\zeta = \frac{1}{r}\frac{E^m}{E^f(1-v)In(\frac{\pi}{4V^f})^{\frac{1}{2}}}, \tag{7}$$

where $L^f$ and $r$ are respectively the length and radius of fibers, and $\nu$ corresponds to the Poisson's ratio of the matrix.

The Young's moduli of the CNT/PMMA composite obtained from the CG model are compared to those calculated from Eq. (5) in Table 2. The Young's modulus of the polymer matrix reinforced by 10-nm long (5, 5) CNTs, calculated by Krenchel's rule of mixtures, increase respectively from 3.38 to 3.70 $GPa$ for the weight fraction of 3 and 5 $wt\%$. The results show percentage differences of 1.1 and 1.9 % at the weight fraction of 3 and 5 $wt\%$, respectively. By further increasing of CNT weight fraction to 10 wt%, the percentage difference increases to 7.3% where the Young's modulus obtained to be equal to 4.85 $GPa$ as presented in Table 2.

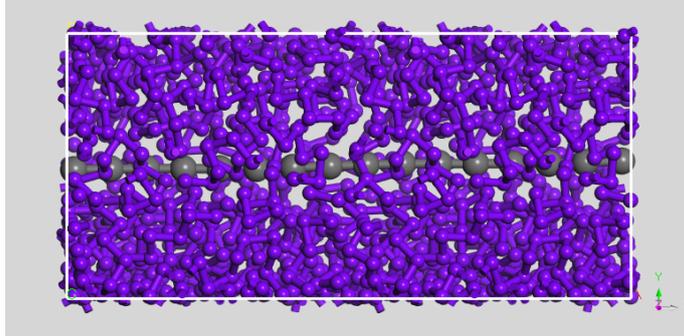

Figure 5: A CG RVE of PMMA polymer matrix with a size of $8 \times 3.7 \times 3.7$ $nm^3$ reinforced by a nanotube rope made of a (5, 5) SWCNT.

To further investigate the verification of the CG model, the Youngs moduli of PMMA polymer and the polymer matrix reinforced by infinite-long (5, 5) CNT obtained from the present CG model are compared to those of MD simulations. For this, a unit cell with a size of $8 \times 3.7 \times 3.7$ $nm^3$ and periodic boundary conditions is constructed as illustrated in Fig. 5. The Youngs moduli of the polymer matrix and infinite-long CNT/PMMA composite measured by the CG model are 2.88 and 47.12 $GPa$, respectively. The predicted stiffness values are in very good agreement with those of 2.86 $GPa$ for pure polymer and 46.73 GPa for infinite-long CNT/PMMA reported in [13] using MD simulations.

It can be concluded that the simulation results of CG model are in a good agreement with those of the Krenchel's rule of mixtures for unidirectional CNT fibers and with those of MD simulation of the CNT/PMMA composites. The micromechanical model is useful for predicting the elastic properties of aligned CNTs reinforced PMMA matrix. However, the development of an accurate micromechanical model for estimating the elastic properties of randomly distributed short fiber composites is quite difficult due to the complicated interactions at the interface of the fibers and matrix, and the complex fiber length and orientation effects. Therefore, the CG model, accounting for the key interactions between polymer matrix and nanotubes, is indispensable for modeling polymer matrix composites reinforced by randomly oriented short CNTs.

*3.2. Elastic Properties of Randomly Distributed CNTs Reinforced PMMA Matrix*

To obtain a statistical representative of the whole micro-structure which results in effective properties of composite materials, a sufficiently large sample volume has to be selected. Based on the work of Hill [35], a representative volume element (RVE) (1) has to be structurally typical of the whole mixture and (2) has to contain an adequate number of inclusions which results in an independent surface values of displacement from the overall moduli. In this study, polymer matrices with a constant thickness of 5 nm in the z direction and variable side lengths in the x and y directions, varying from 20 to 60 nm, are investigated. The polymer matrices are reinforced by 10-nm long CNTs randomly distributed in-plane as shown in Fig 6. The CNT weight fraction and the mass density of the CNT/PMMA composite are set to be 5% and 1.1 $\frac{g}{cm^3}$, respectively. To obtain quasi-isotropic mechanical properties, a uniform probability distribution function is used to place the CNTs inside the polymer matrix.

A suitable RVE size explored by using a successive sample enlargement test. The average Young's modulus of RVEs with side lengths of $l$ and $l'$ are calculated, where $l'$ is larger than $l$ ($l' > l$). The RVE with a side length of $l'$ is taken to be large enough, if the following criterion is satisfied [36],

$$\frac{|E_{l'} - E_l|}{E_{l'}} < 0.01. \quad (8)$$

In Eq. (8), $E_l$ and $E_{l'}$ represent the Young's modulus of an RVE with the size of $l$ and $l'$, respectively.

In order to choose a suitable size for an RVE, the CNT/PMMA composite is simulated with different side lengths. The simulation results, listed in Table 3, demonstrate an increase in the average Young's modulus of the CNT/PMMA composite from 2.85 to 2.97 $GPa$ for the side lengths of 20 and 40 $nm$, respectively showing a tolerance percentage of 4.04%. The average Young's modulus is 2.98 $GPa$ for the side length of 60 $nm$, revealing the percentage difference of 0.34% which meet the criterion introduced in Eq 8. Therefore, the unit cell size of $60 \times 60 \times 5$ $nm^3$ is chosen as the proper RVE. Furthermore, the Young's moduli in the x and y directions of the composite are obtained to be 2.89 and 3.07 GPa at the RVE size as presented in Table 3, indicating a percentage difference less than 6%. It implies that the quasi-isotropic elastic properties are achieved at the sufficiently large RVE with randomly distributed CNT reinforcements. The CG model of the composite system contains 121116 beads that are equivalent to 2065020 atoms in a full atomistic system.

After ascertaining the proper RVE size, the effect of CNT weight fractions on the stiffness of CNT/PMMA composites with randomly distributed fibers is presented in Table 4. The size of simulation box is set to be $60 \times 60 \times 5$ $nm^3$. The length of CNTs and the mass density of the CNT/PMMA composite are the same as previous simulations. As seen in Table 4, with an increase of CNT weight fraction from 0 (pure polymer) to 5 and 8 $wt\%$, the average Young's modulus of the CNT/PMMA composite increases from 2.77 to 2.98 and 3.10 $GPa$, showing percentage increases from 7.05 to 10.6%. By further increasing CNT weight fraction to 10 $wt\%$, the average Young's modulus of the composite is raised to 3.13 $GPa$, revealing a percentage increase of 11.5%. From Table 3, the Young's moduli of a CNT(8 $wt\%$)/PMMA in the x and y directions are measured to be 3.07 and 3.12 $GPa$, respectively. It shows a percentage difference of only 1.6%. The percentage difference increases to 7% at the CNT weight fraction 10%, where the Young's moduli of the CNT/PMMA composite in the x and y directions are respectively 3.02 and 3.25 $GPa$. It can be concluded that although the quasi isotropic elastic properties are attained using the application of randomly distributed short CNT reinforcements, the short fibers do not induce a significant increase in the stiffness of the PMMA polymer matrix. The demanding need for having both quasi isotropic properties and a sufficient level of stiffness comparable to unidirectional CNT reinforced polymer composites has motivated different approaches for enhancing the mechanical properties of polymer composites with randomly distributed reinforcements. In this regard, the formation of cross links between the polymer matrix and nanotubes, and polymer chains is a suitable method for substantially increasing the elastic properties of the composites.

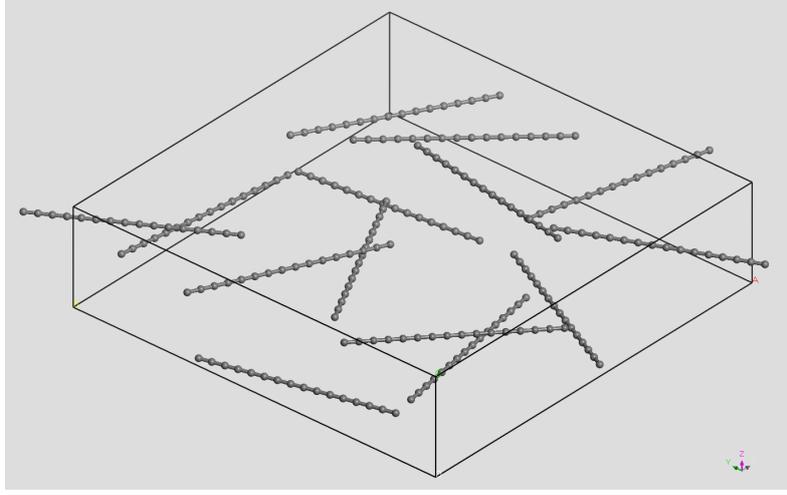

Figure 6: A CNT/PMMA composite with an RVE size of $20 \times 20 \times 5$ $nm^3$ and a CNT weight fraction of 5 $wt\%$. 10-nm long (5, 5) CNTs are randomly distributed in the plane. For a better illustration of the CNT distribution in the composite, polymer chains are omitted.

Table 3: Effect of RVE sizes on Young's modulus of a PMMA matrix reinforced by 10-nm long (5, 5) CNTs randomly distributed in plane. The weight fraction of CNTs is set to be 5 $wt\%$.

| RVE size ($nm^3$) | $E_x(GPa)$ | $E_y(GPa)$ | $E_{av}(GPa)$ | Tolerance (%) |
|---|---|---|---|---|
| $20 \times 20 \times 5$ | 2.79 | 2.92 | 2.85 | - |
| $40 \times 40 \times 5$ | 2.91 | 3.04 | 2.97 | 4.04% |
| $60 \times 60 \times 5$ | 2.89 | 3.07 | 2.98 | 0.34% |

Table 4: Effect of CNT weight fractions on Young's modulus of the CNT/PMMA composite with 10-nm long (5, 5) CNT reinforcement randomly distributed in plane. The RVE size is set to be $60 \times 60 \times 5$ $nm^3$.

| $wt\%$ | $E_x(GPa)$ | $E_y(GPa)$ | $E_{av}(GPa)$ |
|---|---|---|---|
| 0 | 2.78 | 2.75 | 2.77 |
| 5 | 2.89 | 3.07 | 2.98 |
| 8 | 3.07 | 3.12 | 3.10 |
| 10 | 3.02 | 3.25 | 3.13 |

*3.3. Effect of Cross Links on the Elastic Properties of CNT/PMMA Composites*

In order to further study the elastic properties of CNT/PMMA composites, the effect of cross links between the polymer matrix and nanotubes, and polymer chains are investigated in the following simulations. As mentioned in Sec. 2, $2CH_2$ and EGDMA cross links are respectively formed between PMMA polymer and nanotubes, and polymer chains. The RVE size of the CNT/PMMA composite and the weight fraction of CNTs are set to be $60 \times 60 \times 5$ $nm^3$ and 10 $wt\%$, respectively. The length of CNTs are the same as previous simulations.

To find a global minimum energy system, the procedure begins similar to the procedure described in Subsec. 3.1. After the second NPT simulation, $2CH_2$ and EGDMA cross links are added to the composite system. In order to equlibrate the system, a NPT simulation is performed at room temperature of 298 K and atmospheric pressure of 100 kPa in 50 ps with a time step of 1 fs. Another NPT is carried out at the same pressure and temperature conditions for 1 ns with a time step of 10 fs. The NPT simulations are then followed by a further energy minimization to remove internal stresses in the system. The constant-strain minimization method is then applied to the equlibrated system as aforementioned. The procedure continues until the variation of stress versus applied strain is provided to obtain the elastic properties of the composite.

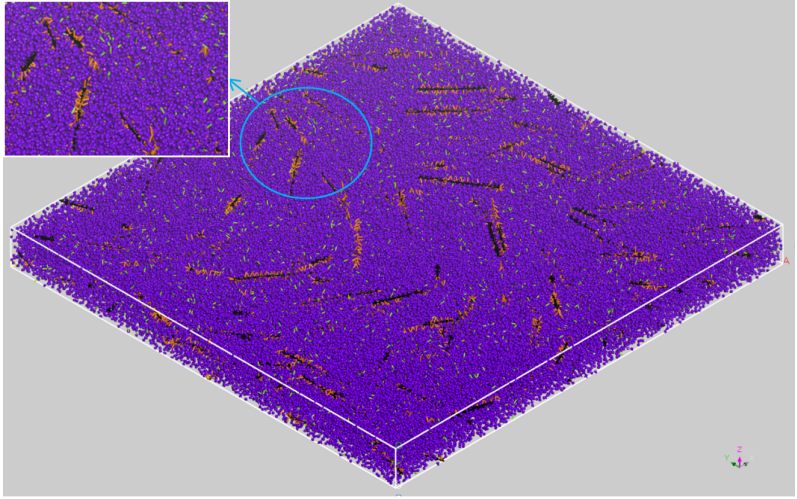

Figure 7: A CNT (10 $wt\%$)/PMMA composite with EGDMA cross links between polymer chains and $2CH_2$ cross links between polymer chains and CNTs. The RVE size is $60 \times 60 \times 5\ nm^3$ and the 10-nm long (5, 5) CNT reinforcements are randomly distributed in plane. Cross links between C-P beads and P-P beads are specified with orange and green lines, respectively. The RVE contains 121120 beads which are equivalent to 2065091 atoms.

The Young's modulus of the CNT (10 $wt\%$)/PMMA composite with $2CH_2$ cross links versus the cross link mole fractions is presented in Fig.8. The cross link mole fraction is defined as the ratio of the amount of mole of cross links to the total amount of moles of the composite system. With an increase in the mole fraction of $2CH_2$ cross links from 0 to 6%, the average Young's modulus of the CNT/PMMA composite increases from 3.13 to 3.63 $GPa$, showing a percentage increase of 16%. By further increasing the mole fraction to 9%, the average Young's modulus of the composite is calculated to be 4.12 $GPa$, which is respectively 32% and 43% stiffer than the composite without cross links and the pure polymer matrix. It can be concluded that the functionalization of CNT reinforcements enables to effectively enhance the elastic properties of the polymer composites.

To further investigate the effect of cross links on the elastic properties of CNT/PMMA composites, the influence of EGDMA cross links between polymer chains on the Young's moduli of the composites are presented in Table 5. In the simulations, the mole fraction of $2CH_2$ cross links between nanotubes and polymer chains is set to be 8%, while the mole fraction of EGDMA cross links varies from 0 to 8%. With an increase in the mole fraction of EGDMA cross links from 0 to 2%, the average Young's modulus of the CNT/PMMA composite is raised from 4.0 to 4.1 $GPa$, showing a percentage increase of 2.5% . The average Young's modulus of the composite with both cross links of $2CH_2$ and EGDMA increases to 4.25 $GPa$ at the mole fractions of 8%, showing a percentage increase of 34.2% and 47.6% compared to the composite without cross links and pure polymer. From the simulation results, the average Young's modulus of the randomly distributed CNT (10 $wt\%$)/PMMA with the cross link mole fractions of 8% (i.e, $E_{av} = 4.25$ $GPa$) is approximately close to the stiffness of an unidirectional CNT (10 $wt\%$)/PMMA composite without cross links (i.e., $E_{av} = 4.52$ $GPa$). It implies that the formation of covalent cross links in the polymer composite notably improves the elastic properties of randomly distributed CNT/PMMA composites. In addition, the Young's moduli of the CNT/PMMA composite with EGDMA mole fraction of 2% and $2CH_2$ mole fraction of 8% are respectively 3.99 and 4.01 $GPa$ in the x and y directions, showing a percentage difference of only 0.5%. By further increasing the mole fraction of EGDMA to 8%, the Young's moduli in the x and y directions are calculated to be 4.29 and 4.20 $GPa$, respectively, revealing a percentage difference up to 2.1%. These results confirm the quasi-isotropic property of the CNT/PMMA composite.

Table 5: Effect of cross links between PMMA chains on Young's modulus of CNT (10 $wt\%$)/PMMA composites with the RVE size of $60 \times 60 \times 5$ $nm^3$ reinforced by 10-nm long (5, 5) CNTs randomly distributed in plane. The mole fraction of the cross links between CNTs and PMMAs is set to be 8%.

| Mole fraction of the polymers cross link % | $E_x(GPa)$ | $E_y(GPa)$ | $E_{av}(GPa)$ |
|---|---|---|---|
| 0 | 3.98 | 4.01 | 4.00 |
| 2 | 3.99 | 4.12 | 4.10 |
| 6 | 4.25 | 4.15 | 4.20 |
| 8 | 4.29 | 4.20 | 4.25 |

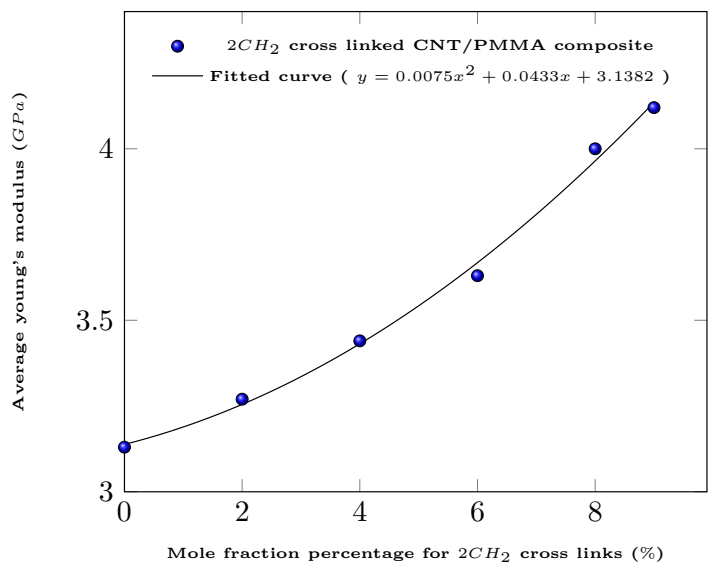

Figure 8: Effect of $2CH_2$ cross links between CNTs and PMMA chains on Young's modulus of CNT (10 $wt\%$)/PMMA composites. The RVE size is set to be $60 \times 60 \times 5$ $nm^3$ and 10-nm long (5, 5) CNTs are randomly distributed in plane.

## 4. Conclusions

The elastic properties of PMMA polymer matrix reinforced by short (5, 5) CNTs are studied using a CG model. The distinguishing feature of the CG model is its ability to model covalent cross links between polymer chains, and nanotube reinforcements and polymer matrix. The CG force field parameters for EGDMA cross links between polymer chains, and $2CH_2$ cross links between CNTs and polymer matrix are derived using MD simulations. The reliability of the CG model in predicting the elastic properties of CNT/polymer composites is examined using the Krenchel's rule of mixtures for unidirectional short fibers. The effects of the mole fraction of cross links, and the weight fraction and distribution of CNTs on the elastic properties of short CNT/PMMA composites are explored. The simulation results demonstrate that although the quasi isotropic elastic properties are achieved using randomly distributed short CNT reinforcements, the short fibers do not induce a significant increase in the stiffness of a polymer matrix in the absence of covalent cross links. In contrast, the formation of covalent cross links between the polymer matrix and randomly distributed nanotubes, and polymer chains is an effective method to attain both high levels of stiffness, which comparable to the stiffness of unidirectional CNT/PMMA composites, and the quasi isotropic elastic properties. From the simulation results, the average Young's modulus of randomly distributed short CNT/PMMA composites with a CNT weight fraction 10 $wt\%$ increases to 4.6 GPa in the presence of $2CH_2$ and EGDMA cross links with a mole fraction of 8%, which is respectively 60 and 47% stiffer than a pure PMMA material and a CNT/PMMA composite with the same nanotube weight fraction. It should be noted that the effect of $2CH_2$ cross links between nanotubes and polymer chains on the enhancement of elastic properties of the composites is more considerable than EGDMA cross links which connect polymer chains.

## 5. Acknowledgments

The authors thank the support of the European Research Council-Consolidator Grant (ERC-CoG) under grant "Computational Modeling and Design of Lithium-ion Batteries (COMBAT)".